\pdfoutput=1

\documentclass[twocolumn,aps,pra,superscriptaddress]{revtex4}

\usepackage{graphicx}
\usepackage{amsfonts}
\usepackage{amsmath}
\usepackage{amssymb}
\usepackage{color}

\newcommand{\ket}[1]{\vert #1 \rangle}
\newcommand{\bra}[1]{\langle#1|}

\def\endproof{\vrule height6pt width6pt depth0pt} 

\begin{document}


\title{Experimental Implementation of a Kochen-Specker Set of Quantum Tests}




\author{Vincenzo~D'Ambrosio}
 \affiliation{Dipartimento di Fisica, ``Sapienza''
 Universit\`{a} di Roma, I-00185 Roma, Italy}

\author{Isabelle~Herbauts}
 \affiliation{Department of Physics, Stockholm University, S-10691
 Stockholm, Sweden}

\author{Elias~Amselem}
 \affiliation{Department of Physics, Stockholm University, S-10691
 Stockholm, Sweden}

\author{Eleonora~Nagali}
 \affiliation{Dipartimento di Fisica, ``Sapienza''
 Universit\`{a} di Roma, I-00185 Roma, Italy}

\author{Mohamed~Bourennane}
 \affiliation{Department of Physics, Stockholm University, S-10691
 Stockholm, Sweden}

\author{Fabio~Sciarrino}
 \affiliation{Dipartimento di Fisica, ``Sapienza''
 Universit\`{a} di Roma, I-00185 Roma, Italy}
 \affiliation{Istituto Nazionale di Ottica (INO-CNR),
 Largo E. Fermi 6, I-50125 Firenze, Italy}

\author{Ad\'an~Cabello}
 \affiliation{Departamento de F\'{\i}sica Aplicada II, Universidad de
 Sevilla, E-41012 Sevilla, Spain}
 \affiliation{Department of Physics, Stockholm University, S-10691
 Stockholm, Sweden}


\date{\today}







\begin{abstract}
The conflict between classical and quantum physics can be identified through a series of yes-no tests on quantum systems, without it being necessary that these systems be in special quantum states. Kochen-Specker (KS) sets of yes-no tests have this property and provide a quantum-versus-classical advantage that is free of the initialization problem that affects some quantum computers. Here, we report the first experimental implementation of a complete KS set that consists of 18 yes-no tests on four-dimensional quantum systems and show how to use the KS set to obtain a state-independent quantum advantage. We first demonstrate the unique power of this KS set for solving a task while avoiding the problem of state initialization. Such a demonstration is done by showing that, for 28~different quantum states encoded in the orbital-angular-momentum and polarization degrees of freedom of single photons, the KS set provides an impossible-to-beat solution. In a second experiment, we generate maximally contextual quantum correlations by performing compatible sequential measurements of the polarization and path of single photons. In this case, state independence is demonstrated for 15~different initial states. Maximum contextuality and state independence follow from the fact that the sequences of measurements project any initial quantum state onto one of the KS set's eigenstates. Our results show that KS sets can be used for quantum-information processing and quantum computation and pave the way for future developments.
\end{abstract}

\maketitle


\section{Introduction}
\label{SecI}


The classical description of nature is based on the assumption that all physical systems possess properties, such as position and velocity, that can be revealed by the act of observation and whose objective existence is independent of whether or not the observation actually does take place. A consequence of this assumption is that a joint probability distribution should exist for the results of any set of joint measurements that reveal these properties \cite{Fine82}. However, there is a fundamental theorem that states that, if quantum mechanics (QM) is correct, then nature cannot be described in classical terms \cite{Specker60,Bell66,KS67}. Kochen and Specker (KS) have provided a particularly appealing proof of this theorem \cite{KS67}, which is valid for systems in any quantum state and which therefore does not require the system to be prepared in specific quantum states, as is the case for the violation of Bell inequalities \cite{Bell64}.

KS have proven that, for any quantum system of dimension $d\ge 3$, there are sets of yes-no tests (represented in QM by projectors $\Pi_i = |v_i\rangle \langle v_i|$ onto unit vectors $|v_i\rangle$) for which it is impossible to assign results 1 (yes) or 0 (no) in agreement with two predictions of QM. (i) If two exclusive tests (represented by orthogonal projectors) are performed on the same system, both cannot give the result~1. (ii) If $d$ pairwise exclusive tests (i.e., satisfying $\sum_{i=1}^d \Pi_i= \mathbb{I}$, with $\mathbb{I}$ the $d$-dimensional identity matrix) are performed on the same system, then one of the tests gives~1. For a given $d$, these sets, called KS sets, are universal in the sense that assigning results is impossible for {\em any} quantum state. The existence of KS sets demonstrates that, for any quantum state, it is impossible to reproduce the predictions of QM with theories in which the measurement results are independent of other compatible measurements. These theories are called noncontextual hidden variable (NCHV) theories.

\begin{figure*}[ht]
\vspace{-7cm}
\centerline{\includegraphics[width=17cm]{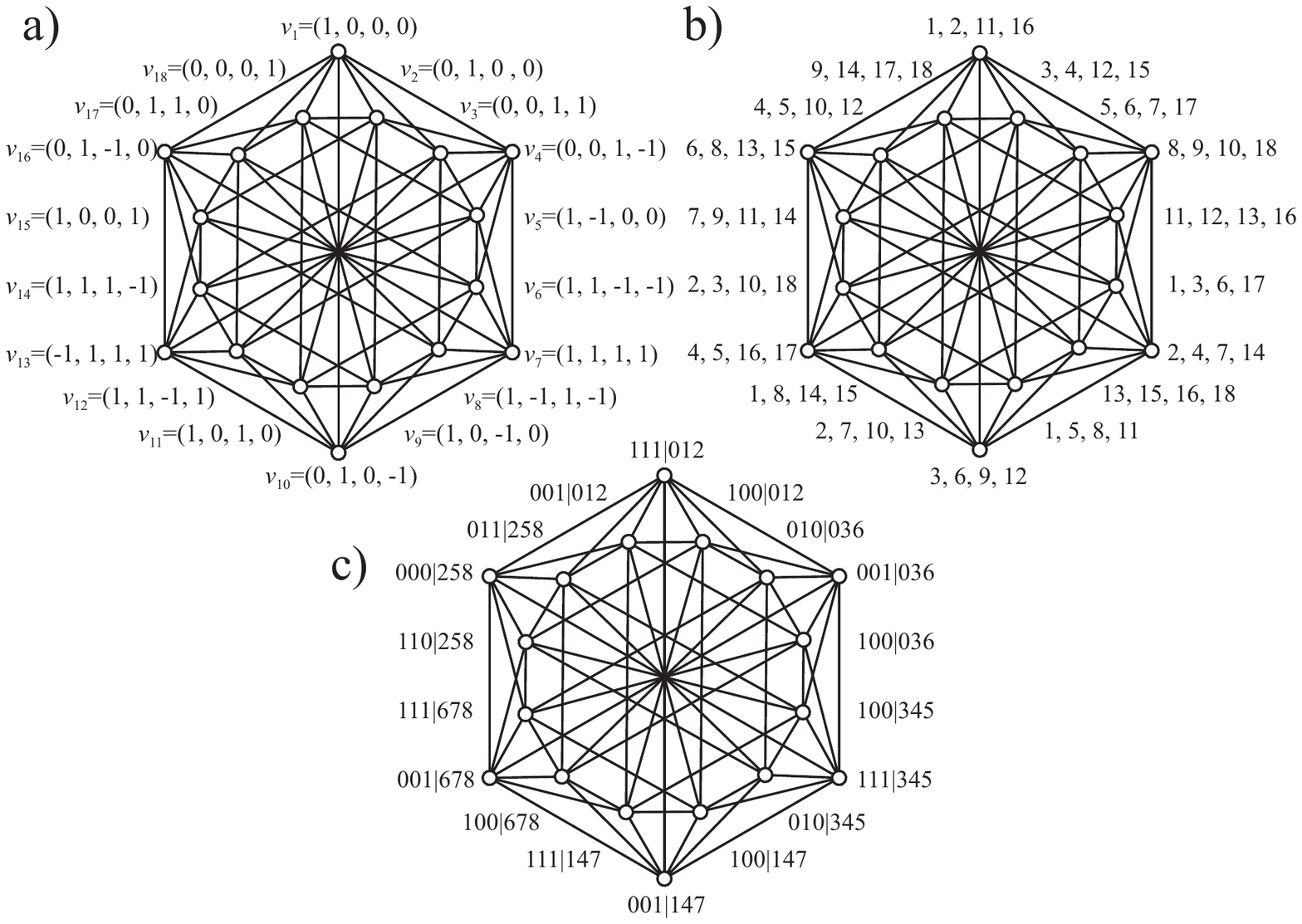}}
\caption{\label{Fig1}
(a)~The 18-test KS set. Each vertex represents a yes-no test (associated in QM with a projector $\Pi_i=|v_i\rangle \langle v_i|$, where
$\langle v_i|$ are the unit vectors displayed in the Figure; normalization factors are omitted to simplify the notation), and adjacent vertices correspond to exclusive tests (i.e., they cannot both have the answer yes on the same system; in QM, they are associated with orthogonal projectors). This vector representation is the one that is adopted in our experiments.
(b)~Optimal strategy for the task that is described in the text using classical resources. The system is a ball that can be placed in one out of 18~boxes, and ``1, 2, 11, 16'' denotes the following yes-no test: ``Is the ball in box 1 or in box 2 or in box 11 or in box 16?.'' The set of classical tests in (b) results in the maximum probability of obtaining yes by using classical resources (see the Appendix~\ref{AppA}).
(c)~Propositions tested in the noncontextuality inequality (\ref{eq:ineq}) that are used to obtain state-independent maximally contextual quantum correlations. Each vertex represents a proposition $abc|xyz$ that denotes that ``the result of measuring $x$ is $a$, the result of measuring $y$ is $b$, and the result of measuring $z$ is $c$.'' When the measurements are those measurements in (\ref{mermin}), then each of these sequences of measurements and results projects any initial state onto the corresponding state in (a).}
\end{figure*}

The original KS set had 117 yes-no tests in $d=3$ \cite{KS67}. In $d=3$, the simplest known KS set has 31 tests \cite{Peres93}, and it has been proven that a KS set with less than 19 tests does not exist \cite{PMMM05,Cabello06,AOW11}. Indeed, numerical evidence suggests that there is no KS set with less than 22 tests in $d=3$ \cite{PMMM05}. However, in $d=4$, there is a KS set with 18 yes-no tests \cite{CEG96}, and it has been proven that there is no KS set with a smaller number of yes-no tests \cite{PMMM05,Cabello06}. Moreover, there is numerical evidence that the same holds for any dimension \cite{PMMM05}, suggesting that, as conjectured by Peres \cite{Peres03}, the 18-test KS set is the simplest one in any dimension. A graph can be associated with any KS set \cite{KS67}. In this graph, each yes-no test of the KS set is represented by a vertex and exclusive yes-no tests are represented by adjacent vertices. Figure \ref{Fig1}(a) shows the graph corresponding to the 18-test KS set. Other proofs of state-independent quantum contextuality based on observables represented by Pauli operators \cite{Peres90,Mermin90} can be expressed in terms of KS sets by noticing that the projectors onto the common eigenstates of the commuting Pauli operators constitute a KS set \cite{Peres91,KP95}. Some recent proofs of state-independent quantum contextuality are not based on KS sets but on sets of yes-no tests for which an assignment satisfying (i) and (ii) exists \cite{YO12,BBC12}. The necessary condition for state-independent quantum contextuality, common to KS sets and these new sets, is described in \cite{Cabello12b}. One of these new sets has only 13 yes-no tests \cite{YO12}. However, the orthogonality graph (i.e., the one constructed taking all vectors with equal weight) corresponds to an inequality without quantum violation; the quantum violation requires that nine of the vectors appear with double weight, so the corresponding (unweighted) graph has 22 vertices. The same holds true for the graph associated with the corresponding tight inequality \cite{KBLGC12}.

While Bell inequalities \cite{Bell64} that reveal quantum nonlocality have stimulated a large number of experiments (e.g., \cite{ADR82,TBZG98,WJSWZ98,RMSIMW01}) and have a number of applications (e.g., \cite{Ekert91,BZPZ04,PAMBMMOHLMM10}), the awareness that quantum contextuality and, specifically, state-independent quantum contextuality can also be observed in actual experiments is relatively recent \cite{Cabello08}. On one hand, there are quantum-contextuality experiments with photons \cite{Michler00, Lapkiewicz11, ADLPBC12} and neutrons \cite{Bartosik09}, in which the system has to be prepared in a special state. On the other hand, the state-independent quantum-contextuality experiments with ions \cite{Kirchmair09}, photons \cite{Amselem09}, and nuclear-magnetic-resonance systems \cite{Moussa10} test the violation of a noncontextuality inequality that involves observables represented by Pauli operators. A complete KS set of yes-no tests, in the original form defined by KS, has never been experimentally implemented. As mentioned before, the 13 yes-no tests in \cite{YO12} are not a KS set (although they belong to a KS set of 33 yes-no tests \cite{Peres91}). Therefore, the experiment in \cite{ZWDCLHYD12} cannot be considered an implementation of a KS set. Moreover, it can hardly be considered an experiment of contextuality, since each test is performed using a different device, depending on the context \cite{ABBCGKLW12}. A proper way to carry out the experiment has been proposed in \cite{CABBB12}.

In this paper, we present the first experimental implementation of a KS set of yes-no tests. We report the results of two experiments. In the first one, described in Sec.~\ref{SecII}, we use the polarization and orbital angular momentum of single photons to show how a KS set can be used to obtain a state-independent impossible-to-beat quantum-versus-classical advantage in a specific task.

In the second experiment, described in Sec.~\ref{SecIII}, we perform sequential measurements of compatible observables encoded in the path and polarization degrees of freedom of single photons. From the measurements, we then produce correlations that violate a noncontextuality inequality that is constructed in a one-to-one correspondence with the eigenstates of the same KS set. This experiment shows how KS sets can be used to obtain state-independent maximally contextual quantum correlations.

Finally, in Sec.~\ref{SecIV}, we connect both experiments, present the conclusions, and describe near-future applications and further developments that could be pursued in the future.


\section{Experimental observation of state-independent impossible-to-beat KS-based quantum advantage using polarization and orbital angular momentum of photons}
\label{SecII}


Consider the following task \cite{NDSC12}: Given an $n$-vertex graph $G$, provide $n$ yes-no tests about a physical system, such that each test is associated with a vertex of $G$, exclusive tests correspond to adjacent vertices, and these tests result in the highest probability of obtaining a yes answer when one of them is chosen at random. This highest probability may be different, depending on whether the physical system and the tests are classical, quantum, or postquantum. Moreover, for arbitrary graphs, the highest probability may also depend on the state in which the system is prepared. However, two distinguishing features of the graph of Fig.~\ref{Fig1}(a) are that the highest probability in QM can be reached regardless of the state of the system and that such a probability cannot be outperformed by any postquantum theory (see the Appendix~\ref{AppB}).

If the available resources are classical, i.e., physical systems with preassigned results and tests thereof, then an optimal strategy is illustrated in Fig.~\ref{Fig1}(b). There, the classical system is assumed to be a ball that can be placed in one out of 18 boxes numbered from 1 to 18. For instance, ``1, 2, 11, 16'' denotes the following yes-no test: ``Is the ball in box~1 or in box~2 or in box~11 or in box~16?.'' The other tests are shown in Fig.~\ref{Fig1}(b). The 18 tests satisfy the graph's relations of exclusivity. In addition, no matter which box the ball is placed in, the probability of getting a yes answer when one of the 18 tests is chosen at random is $4/18 \approx 0.22$, since the answer is always ``yes'' for 4 of the tests and ``no'' for the others. Alternatively, the performance can be measured by the sum $\Sigma$ of the probabilities of obtaining a yes answer. It can be proven that, for this graph, no other set of classical yes-no tests allows a higher probability (see the Appendix~\ref{AppB}). Therefore, using classical resources,
\begin{equation}
 \label{sigma}
 \Sigma = \sum_{i \in V(G)} P(\Pi_i=1) \le 4,
\end{equation}
where $V(G)$ is the set of vertices of the graph in Fig.~\ref{Fig1}(a) and $P(\Pi_i=1)$ is the probability of obtaining the result $1$ (yes) for the yes-no test $\Pi_i$.

However, it can be easily checked that, if we use the 18 quantum yes-no tests $\Pi_i=|v_i\rangle \langle v_i|$ in Fig.~\ref{Fig1}(a), then the probability of a yes answer is $1/4 = 0.25$ and
\begin{equation}
 \label{qsigma}
 \Sigma_{\rm QM} = 4.5.
\end{equation}
Since this advantage is independent of the initial quantum state of the system, this task is an example of a task with a quantum advantage for which the initialization problem affecting nuclear-magnetic-resonance quantum computers \cite{GC97,CFH97} is not an obstacle. Moreover, for this task, even hypothetical postquantum theories cannot outperform QM (see the Appendix~\ref{AppC}).

In order to test this state-independent impossible-to-beat quantum advantage in an experiment, we use for the encoding 2 different degrees of freedom of the same photon: the polarization and a bidimensional subset of the orbital-angular-momentum space \cite{Naga10pra}, spanned by the states with eigenvalues $m=\pm2 \hbar$. The four-dimensional logical basis for encoding is
\begin{equation}
 \{\ket{H, +2}, \ket{H, -2}, \ket{V, +2}, \ket{V, -2}\},
 \label{basisRoma}
\end{equation}
where $H$ and $V$ denote horizontal and vertical polarization, respectively, and $\pm 2$ denotes $m=\pm2 \hbar$.

\begin{figure*}[ht]
\vspace{5cm}
\centerline{\includegraphics[scale=0.46]{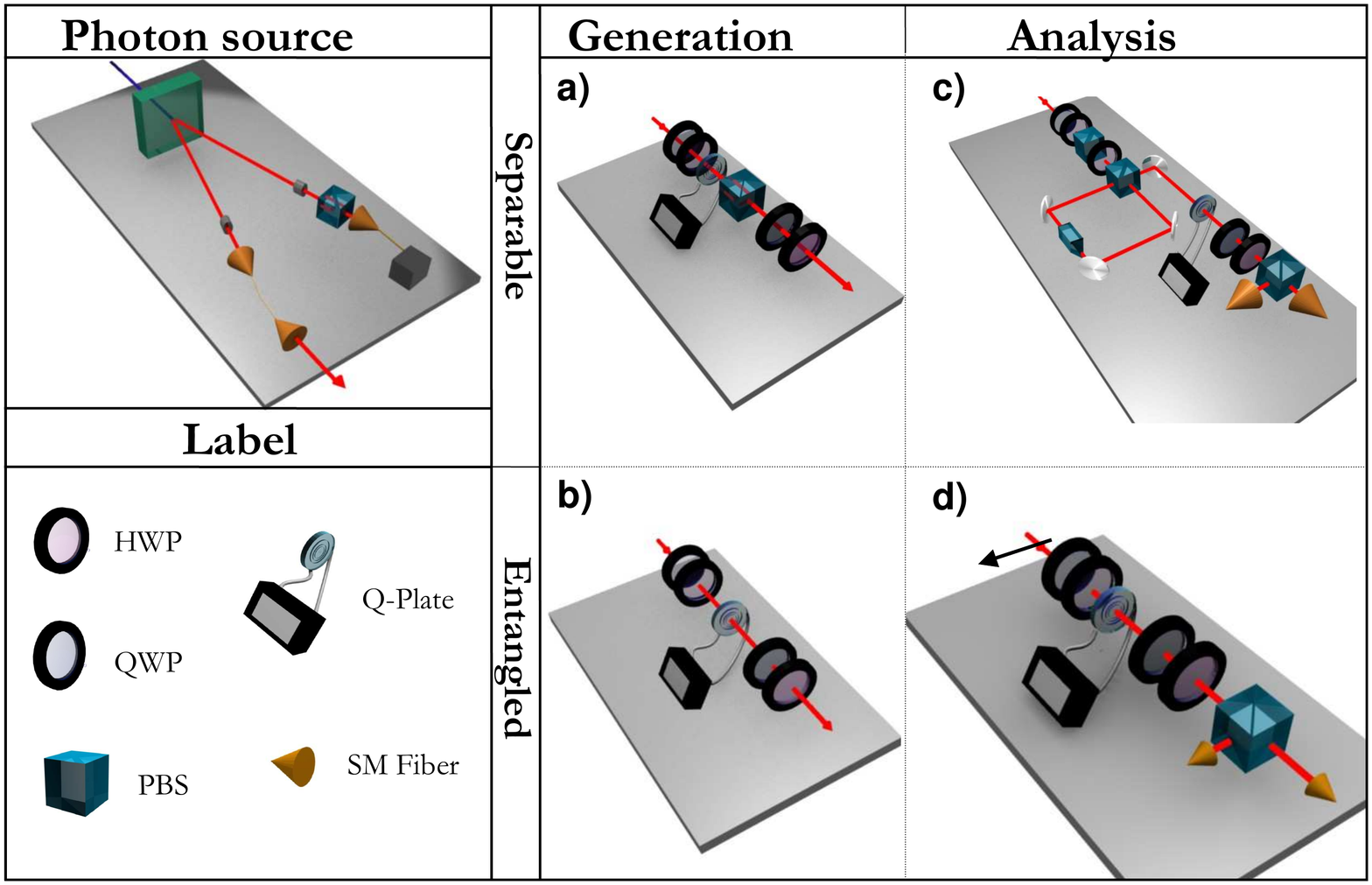}}
\caption{\label{Fig2} Experimental setup for the measurement of the probabilities $P(\Pi_i=1)$ on different states encoded in the space of polarization and orbital-angular-momentum. In the upper left corner, the single-photon source is represented. The four schemes we use for the experiment are presented in the right part of the figure. Each state is prepared by one of the two setups of the column labeled ``Generation'': setup (a) for separable states (quantum transferrer $\pi\rightarrow o_2$ \cite{Naga09opt}) and setup (b) for entangled ones (an ``entangler'' based on a QP and waveplates). The column labeled ``Analysis'' shows the setups for the projection onto the desired states: setup (c) for separable states, a deterministic transferrer $o_2\rightarrow \pi$, and setup (d) for entangled states, where a QP is needed to have a deterministic detection.}
\end{figure*}

The experimental setup involves preparing the required states (preparation stage) and then projecting them onto the desired states (measurement stage).
In the preparation stage, heralded single-photons of 795-nm wavelength are produced in a noncollinear parametric down-conversion process where a beta-barium-borate crystal is pumped by the second harmonic of a pulsed laser with a repetition rate of 76 MHz. The single photons are then coupled to a single-mode (SM) fiber in order to filter out all the transverse electromagnetic modes but the fundamental TEM$_{00}$ one (which is an orbital-angular-momentum eigenstate with eigenvalue equal to zero). The second photon generated in the spontaneous parametric down conversion acts as a trigger of the single-photon generation. After the SM fiber, the input photon is prepared using half-wave plates (HWPs), quarter-wave plates (QWPs), $q$ plates (QPs), and polarizing beam splitters (PBSs) to generate the required states in the logical basis (\ref{basisRoma}). As explained in Fig.~\ref{Fig2}, the procedure is different depending on whether the state to be generated is separable or entangled.

The QPs are liquid-crystal devices that produce a spin-orbit coupling of the polarization and orbital-angular-momentum contributions to the photons' total angular momentum \cite{Marr11}. When a photon interacts with the QP, it suffers an exchange of orbital angular momentum that is driven by the input polarization. In particular, for the QPs adopted in this experiment, the shift of orbital angular momentum is equal to $\pm2\hbar$ when the input photon has left (right) polarization \cite{Naga09prl,Naga09opt}. The QP efficiency has been optimized by controlling the electrical tuning \cite{Picc10apl}, leading to a conversion efficiency of $94\%$. Thanks to its features, the QP can be adopted for both the generation and the analysis of quantum states encoded in the orbital angular momentum.


The measurement stage is achieved by using a deterministic transferrer based on a Sagnac interferometer, with a Dove prism in one of the arms when the prepared state is separable \cite{Damb12} and with a QP with a standard polarization-analysis setup when the state is entangled. For each state to be analyzed, we record the coincidence counts between the trigger and the signal coupled through the SM fiber at the end of the measurement setup. Considering all loss contributions in the setup, we record around $30$ Hz as mean coincidence counts. The experimental results for $\Sigma$, as measured on 15 different states, are reported in Table~\ref{Romatable}. The experimental data are in good agreement with the theoretical prediction, with a mean value of $\Sigma_{\rm exp}= 4.512 \pm 0.005$ to be compared to $\Sigma=4.5$, and show the clear advantage of the quantum settings with KS projectors over any classical strategy.


\begin{table}[ht]
{\small
\begin{tabular}{|c|c|c|c|}
\hline
\hline
\textbf{Code} & \textbf{State} & \textbf{Implementation} & \textbf{$\Sigma$} \\
\hline\hline
$v_1$ & (1,0,0,0) & $\ket{H,+2}$ & $4.60 \pm0.02$ \\ \hline
$v_2$ & (0,1,0,0) & $\ket{H,-2}$ & $4.45 \pm0.02$\\ \hline
$v_7$ & (1,1,1,1) & $\ket{A,h}$ & $4.50 \pm0.02$\\ \hline
$v_{11}$ & (1,0,1,0) & $\ket{A,+2}$ & $4.51 \pm0.02$\\ \hline
$v_{15}$ & (1,0,0,1) & $\ket{\psi_1}=\frac{1}{\sqrt{2}}(\ket{H,+2}+\ket{V,-2})$ & $4.59 \pm0.02$\\ \hline
$v_{16}$ & (0,1,-1,0) & $\ket{\psi_3}=\frac{1}{\sqrt{2}}(\ket{H,-2}-\ket{V,+2})$ & $4.47 \pm0.01$\\ \hline
$v_{17}$ & (0,1,1,0) & $\ket{\psi_4}=\frac{1}{\sqrt{2}}(\ket{H,-2}+\ket{V,+2})$ & $4.41 \pm0.02$\\ \hline
$v_{18}$ & (0,0,0,1) & $\ket{V,-2}$ & $4.50 \pm0.02$\\ \hline
$v_{19}$ & (0,0,1,0) & $\ket{V,+2}$ & $4.45 \pm0.03$\\ \hline
$v_{20}$ & (1,1,0,0) & $\ket{H,h}$ & $4.57 \pm0.02$\\ \hline
$v_{24}$ & (1,0,0,-1) & $\ket{\psi_2}=\frac{1}{\sqrt{2}}(\ket{H,+2}-\ket{V,-2})$ & $4.58 \pm0.02$\\ \hline
$\rho_{25}$ && $\frac{13}{16}\ket{\psi_1}\bra{\psi_1}+\frac{1}{16}\sum_{j=2}^4\ket{\psi_j}\bra{\psi_j}$ & $4.57\pm0.02$\\ \hline
$\rho_{26}$ && $\frac{5}{8}\ket{\psi_1}\bra{\psi_1}+\frac{1}{8}\sum_{j=2}^4\ket{\psi_j}\bra{\psi_j}$ & $4.55\pm0.02$\\ \hline
$\rho_{27}$ && $\frac{7}{16}\ket{\psi_1}\bra{\psi_1}+\frac{3}{16}\sum_{j=2}^4\ket{\psi_j}\bra{\psi_j}$ & $4.53\pm0.02$\\ \hline
$\rho_{28}$ && $\frac{1}{4}\sum_{j=1}^4\ket{\psi_j}\bra{\psi_j}$ & $4.50\pm0.02$\\ \hline
 \hline
 \multicolumn{3}{|c|}{\textbf{Average value}} & \textbf{$4.512 \pm 0.005$}\\
\hline
\end{tabular}
}
\caption{\label{Romatable}Experimental results for $\Sigma$ for 15 quantum states. Each input state is projected onto each of the 18 states in Fig.~\ref{Fig1}(a). Notation for this Table includes $\ket{A} = \frac{\ket{H}+\ket{V}}{\sqrt{2}}$ and $\ket{h} = \frac{\ket{+2}+\ket{-2}}{\sqrt{2}}$. The error bars are evaluated by considering the Poissonian statistics of coincidence counts. All reported values lie in the range $[\Sigma_{\rm min},\Sigma_{\rm max}]$ (see Fig.~\ref{Fig3}).}
\end{table}


In addition, the exclusivity between the tests in Fig.~\ref{Fig1}(a) is experimentally verified, confirming that tests corresponding to adjacent vertices cannot both be simultaneously true. Experimentally, the probabilities $P_{|v_j\rangle}(\Pi_i=1)$, obtained by projecting the state $|v_i\rangle$ onto the state $|v_j\rangle$ for orthogonal states (adjacent vertices), are close to 0, as expected. Specifically, we obtain that the mean value of $P_{|v_j\rangle}(\Pi_i=1)$ is $\epsilon=(0.014 \pm 0.001)$ (see Table~V).


\begin{figure*}[ht]
\vspace{-5cm}
\centerline{\includegraphics[scale=0.65]{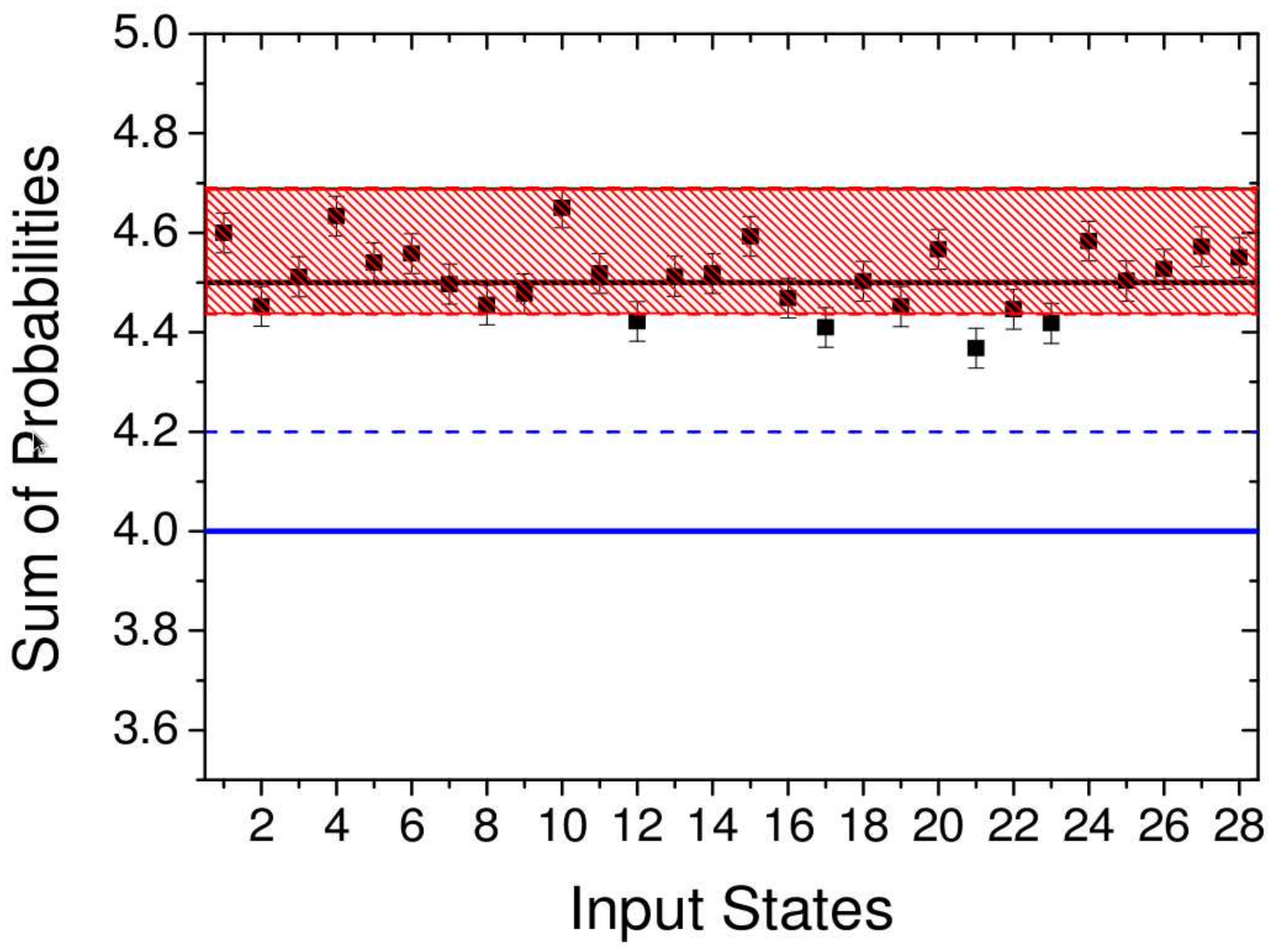}}
\caption{Experimental results for $\Sigma$ for 28 quantum states. The solid (dashed) blue line refers to the (corrected) classical upper bound for $\Sigma$. The red area represents the range $[\Sigma_{\rm min}, \Sigma_{\rm max}]$ in which we theoretically expect to find all experimental values of $\Sigma$. The first 18 states correspond to the ones in Fig.~\ref{Fig1}(a). States 19--28 are defined in Table~\ref{Romatable}.}
\label{Fig3}
\end{figure*}


The theoretical classical and quantum bounds for the task should be properly corrected to take into account that $\epsilon \neq 0$. Assuming that inequality (\ref{sigma}) is only valid with probability $1-\epsilon$ and that the worst-case scenario, in which there are no edges and thus the upper bound of the inequality is 18, occurs with probability $\epsilon$, to certify the quantum advantage it is enough that $4 (1-\epsilon) + 18 \epsilon < \Sigma$, which, using $\Sigma=4.5$, implies $\epsilon <0.035$, a condition that is fulfilled in our experiment. Moreover, we expect to observe a quantum advantage that lies in a range $[\Sigma_{\rm min}, \Sigma_{\rm max}]$, where $\Sigma_{\rm min}=4.5(1-\epsilon)$ and $\Sigma_{\rm max}=4.5(1-\epsilon)+18\epsilon$. Here, $\Sigma_{\rm max}$ ($\Sigma_{\rm min}$) corresponds to the situation of having all 18 propositions proven true (false) with probability $\epsilon$. In Fig.~\ref{Fig3}, we report the experimental values of $\Sigma$, not only for the 15 states in Table~\ref{Romatable} but also for the other 13 states. The quantum advantage is observed for all 28 states, in good agreement with the state-independent value predicted by the theory.


\section{Experimental state-independent maximally contextual quantum correlations by sequential measurements on polarization and path of photons}
\label{SecIII}


KS sets can also be used to generate nonclassical contextual correlations by performing sequential compatible measurements on individual systems. The signature of nonclassicality is the violation of a noncontextuality inequality, which is an inequality involving linear combinations of joint probabilities of sequential compatible measurements, satisfied by any NCHV theory.

For most of the experimental demonstrations of contextual correlations to date \cite{Michler00, Bartosik09, Lapkiewicz11, ADLPBC12}, the system has to be prepared in a special state. There are also theoretical \cite{Cabello08} and experimental works \cite{Kirchmair09, Amselem09, Moussa10} on state-independent contextuality. However, none of the previous experiments implement a KS set of yes-no tests.

Here, we use the KS set of Fig.~\ref{Fig1}(a) to obtain a noncontextuality inequality violated by any quantum state. This inequality follows from identifying sequential compatible measurements such that any initial state is projected onto one of the eigenstates of the yes-no tests of the KS set of Fig.~\ref{Fig1}(a). This correspondence guarantees that the propositions $abc|xyz$ keep all the relations of exclusivity existing in Fig.~\ref{Fig1}(a). [The proposition $abc|xyz$ denotes ``the result of measuring $x$ (first measurement of the sequence) is $a$, the result of measuring $y$ (second) is $b$, and the result of measuring $z$ (third) is $c$.''].

A one-to-one correspondence between the 18 propositions in Fig.~\ref{Fig1}(c) and the 18 states in Fig.~\ref{Fig1}(a) can be established by assigning the results $0$ and $1$ to the degenerate eigenvalues $-1$ and $1$ of the following operators,
\begin{equation}
\label{mermin}
\begin{split}
&0:=\sigma_z \otimes \mathbb{I}, \;\;\;\;\;\;\;\; 1:=\mathbb{I}\otimes \sigma_z,\;\;\;\;\;\;\;\; 2:=\sigma_z \otimes \sigma_z,\\
&3:=\mathbb{I} \otimes \sigma_x , \;\;\;\;\;\;\;\; 4:=\sigma_x\otimes \mathbb{I}, \;\;\;\;\;\;\;\; 5:=\sigma_x \otimes \sigma_x,\\
&6:=\sigma_z \otimes \sigma_x, \;\;\;\;\;\; 7:=\sigma_x\otimes \sigma_z, \;\;\;\;\; 8:=\sigma_y \otimes \sigma_y,
\end{split}
\end{equation}
where $\sigma_x$, $\sigma_y$, and $\sigma_z$ are the Pauli matrices along the $x$, $y$, and $z$ directions and $\otimes$ denotes tensor product. Therefore, the corresponding noncontextuality inequality reads
\begin{equation}
\label{eq:ineq}
\begin{split}
\xi=&P(001|012)+P(111|012)+P(100|012)\\
&+P(010|036)+P(001|036)+P(100|036)\\
&+P(100|345)+P(111|345)+P(010|345)\\
&+P(100|147)+P(001|147)+P(111|147)\\
&+P(100|678)+P(001|678)+P(111|678)\\
&+P(110|258)+P(000|258)+P(011|258)\stackrel{\mbox{\tiny{ NCHV}}}{\leq} 4,
\end{split}
\end{equation}
where the upper bound for NCHV theories follows from the classical bound of inequality (\ref{sigma}). For any initial state, these sequences of quantum measurements lead to
\begin{equation}
 \xi_{\rm QM}=4.5,
\end{equation}
in correspondence with the quantum advantage (\ref{qsigma}). It can be proven that the contextuality revealed by this violation cannot be outperformed by any post-quantum theory (see the Appendix~\ref{AppC}).


\begin{figure*}[t]
\vspace{3cm}
\centerline{\includegraphics[scale=0.58]{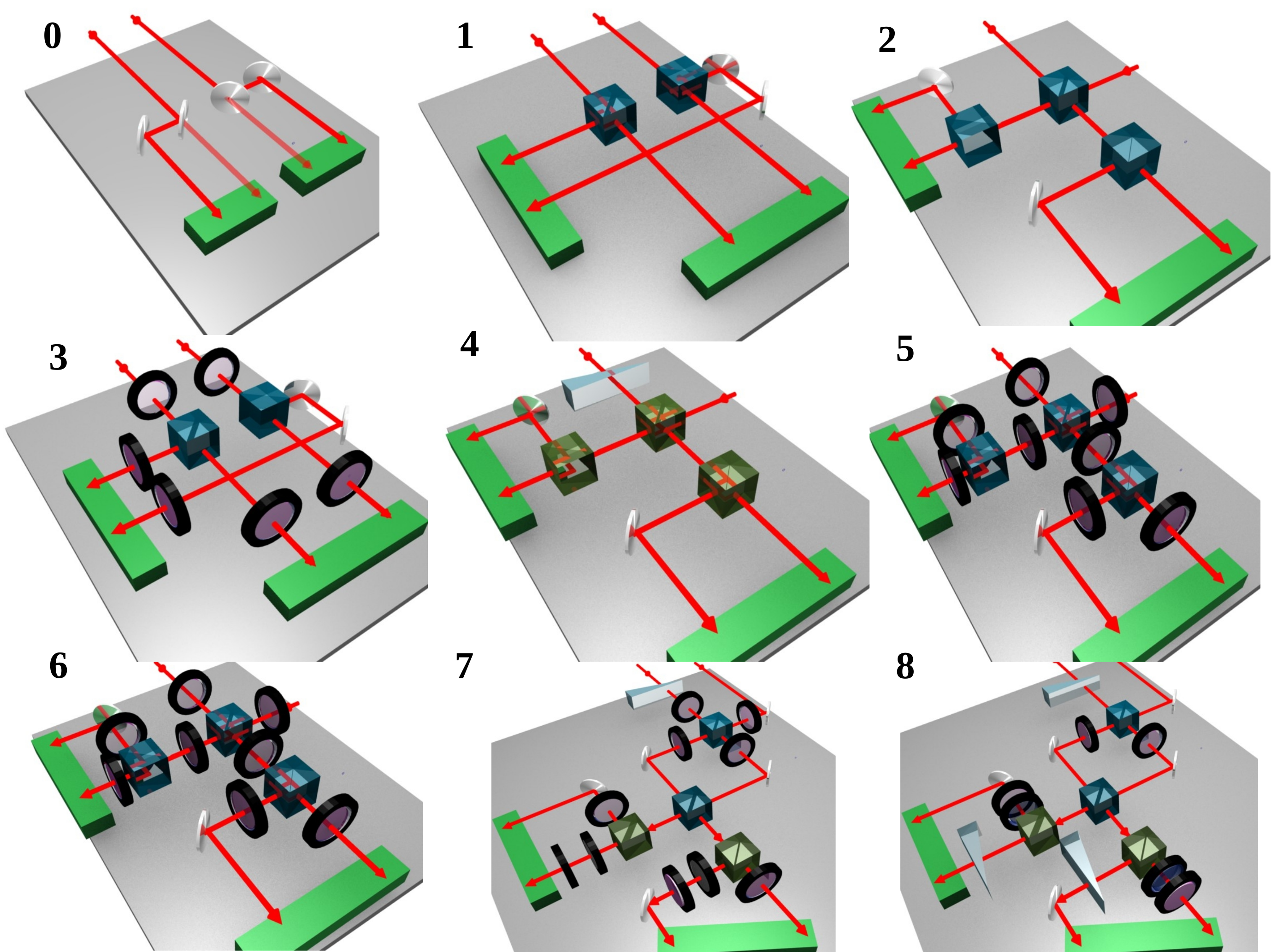}}
\caption{\label{Fig4}Experimental setups for the observables in (\ref{mermin}). For the measurement of observable 0, it is only necessary to distinguish between the paths $r$ and $t$. To measure observable 4, a polarization-independent beam splitter is used to distinguish the eigenstates through interference. The measurements of observables 1 and 3 are standard polarization measurements using PBSs and HWPs. Observables 2, 5, and 8 are Bell-state measurements, and so are the measurements of 6 and 7, but in the latter group the Bell measurement is preceded by a rotation of the polarization to guarantee compatibility with observable 8. To measure the probabilities that appear in inequality (\ref{eq:ineq}), these measurement devices are arranged in a cascaded manner \cite{ADLPBC12, Amselem09}.}
\end{figure*}


We tested inequality (\ref{eq:ineq}) in a separate experiment using a single four-dimensional system with two qubits encoded in the spatial path and two qubits encoded in the polarization of the photon. For this experiment, the logical basis is
\begin{equation}
 \label{basisStock}
 \{\ket{t, H}, \ket{t, V}, \ket{r, H}, \ket{r, V}\},
\end{equation}
where $t$ and $r$ denote the transmitted and reflected paths of the photon, respectively, and $H$ and $V$ denote horizontal and vertical polarization, respectively.

The experiment involves testing a sequence of three compatible measurements that correspond to rows or columns in (\ref{mermin}). To do so, the experimental setup is designed as a cascade of measurement boxes that represent the compatible observables, which is preceded by a preparation device and followed by detectors \cite{ADLPBC12, Amselem09}. The preparation device consists of a source of $H$-polarized single photons that is implemented using a narrow-bandwidth cw diode laser at $780$ nm of long coherence length that is attenuated to a mean photon number of $0.06$ photons per coincidence gate. Combinations of HWPs, PBSs, and a wedge placed after the single-photon source create any desired state in the logical basis \ref{basisStock}. The detection stage uses calibrated silicon avalanche photodiodes, with an eight-channel coincidence logic and a coincidence window of $1.7$ ns.

Crucial for the experimental test of the noncontextuality inequality (\ref{eq:ineq}) is the proper design of the devices for measuring the observables in (\ref{mermin}). These devices should satisfy two conditions. The first condition is compatibility. [The three measurements corresponding to rows and columns in (\ref{mermin}) should be compatible, so that any subsequent measurement of any of them would give the same result.] The second condition is noncontextuality. [Every observable in (\ref{mermin}) has to be measured using the same device in any of the sequences.] These conditions are achieved with the design of the nine measuring devices shown in Fig.~\ref{Fig4}.

To construct the cascade setup, we use displaced Sagnac interferometers with very high stability. We obtain visibilities in the $90$\%--$99$\% range, depending on the implemented sequence. The detection efficiency of the single-photon detectors is $55\%$, and the efficiency of the fiber coupling is $90\%$. The experimental value of $\xi$ for 15 different quantum states is reported in Table~II. Under the assumption that the detected photons are an unbiased subset of the emitted photons (a fair sampling assumption), the results in Table~II are in good agreement with a state-independent violation of inequality (\ref{eq:ineq}). The deviations from the quantum prediction for an ideal experiment with perfect compatibility are due to the systematic errors that arise from the interferometers, the light-mode overlapping, and the imperfection of the polarization components.


\begin{table}[ht]
{\small
\begin{tabular}{|c|c|c|c|}
\hline
\hline
\textbf{Code} & \textbf{State} & \textbf{Implementation} & $\xi$ \\
\hline\hline
 $v_1$ & (1,0,0,0) & $\ket{t, H}$ & $4.1953 \pm 0.0015$ \\ \hline
 $v_2$ & (0,1,0,0) & $\ket{t, V}$ & $4.2690 \pm 0.0025$\\ \hline
 $v_7$ & (1,1,1,1) & $\ket{p, D}$ & $4.3790 \pm 0.0011$\\ \hline
$v_{11}$ & (1,0,1,0) & $\ket{p, H}$ & $4.4406 \pm 0.0024$\\ \hline
$v_{15}$ & (1,0,0,1) & $\ket{\psi_1}=\frac{1}{\sqrt{2}}(\ket{t, H}+\ket{r, V})$ & $4.2607 \pm 0.0011$\\ \hline
$v_{16}$ & (0,1,-1,0) & $\ket{\psi_3}=\frac{1}{\sqrt{2}}(\ket{r, H}-\ket{t, V})$ & $ 4.2550 \pm 0.0020$\\ \hline
$v_{17}$ & (0,1,1,0) & $\ket{\psi_4}=\frac{1}{\sqrt{2}}(\ket{r, H}+\ket{t, V})$ & $ 4.1990 \pm 0.0022$\\ \hline
$v_{18}$ & (0,0,0,1) & $\ket{r, V}$ & $4.3001 \pm 0.0017$\\ \hline
$v_{19}$ & (0,0,1,0) & $\ket{r, H}$ & $4.3346 \pm 0.0030$\\ \hline
$v_{20}$ & (1,1,0,0) & $\ket{t, D}$ & $4.4113 \pm 0.0013$\\ \hline
$v_{24}$ & (1,0,0,-1) & $\ket{\psi_2}=\frac{1}{\sqrt{2}}(\ket{t, H}-\ket{r, V})$ & $4.2468 \pm 0.0011$\\ \hline
$\rho_{25}$ && $\frac{13}{16}\ket{\psi_1}\bra{\psi_1}+\frac{1}{16}\sum_{j=2}^4\ket{\psi_j}\bra{\psi_j}$ & $4.3136 \pm 0.0819$\\ \hline
$\rho_{26}$ && $\frac{5}{8}\ket{\psi_1}\bra{\psi_1}+\frac{1}{8}\sum_{j=2}^4\ket{\psi_j}\bra{\psi_j}$ & $4.3479 \pm 0.0984$\\ \hline
$\rho_{27}$ && $\frac{7}{16}\ket{\psi_1}\bra{\psi_1}+\frac{3}{16}\sum_{j=2}^4\ket{\psi_j}\bra{\psi_j}$ & $4.3171 \pm 0.1080$\\ \hline
$\rho_{28}$ && $\frac{1}{4}\sum_{j=1}^4\ket{\psi_j}\bra{\psi_j}$ & $4.3968 \pm 0.1098$\\
\hline
\end{tabular}
}
\label{Stockholmtable}
\caption{Experimental results for $\xi$ for 15 quantum states. Notation for this Table includes
$\ket{p} = \frac{\ket{t}+\ket{r}}{\sqrt{2}}$ and $\ket{D} = \frac{\ket{H}+\ket{V}}{\sqrt{2}}$.
The errors in the results of $\xi$ are deduced from the standard deviation of 50 samples in the 10-s time period. [See Tables VI--VIII for all the experimental values of $P(abc|xyz)$ for the 15~states].}
\end{table}


\section{Conclusions}
\label{SecIV}


We have presented the first experimental implementation of a KS set of quantum yes-no tests in two experiments using two different four-dimensional photonic systems and with two different purposes.

In the first experiment, we wanted to show the state-independent impossible-to-beat quantum-versus-classical advantage of using the KS set for the specific task described in Sec.~\ref{SecII}. This advantage is observed through the violation of inequality \eqref{sigma}, which holds for any classical implementation of the task. Crucial in this first experiment, since it is crucial for the validity of inequality \eqref{sigma}, is to demonstrate that the conditions defining the task are actually satisfied. In particular, it is crucial to demonstrate that all 42 exclusivity relations between pairs of tests assumed in the definition of the task are actually satisfied in the experiment. For this purpose, the polarization and orbital angular momentum of single photons are ideal: They allow us to perform not only the tests needed to observe the violation of inequality \eqref{sigma} (see Table~V) but also the high number of tests needed to confirm the conditions under which inequality \eqref{sigma} is valid (see Table~V).

In the second experiment, we wanted to produce correlations violating the noncontextuality inequality \eqref{eq:ineq} that is constructed in a one-to-one correspondence with the eigenstates of the KS set to show how KS sets can be used to reveal state-independent maximally contextual quantum correlations. The ability to perform sequential measurements of compatible observables on the same system is crucial for this second test. To achieve this goal, we adopt a different encoding. By using the path and polarization degrees of freedom of single photons we can implement a cascade setup that allows us to perform sequential measurements that guarante, at the same time, that the other fundamental requirement in any test of a noncontextuality inequality is satisfied, namely, that the same observable is measured with the same device in any context.

Our results pave the way for further developments. Near-future applications of our experiments include specific cryptographic applications \cite{Svozil10,CDNS11} and dimension witnessing \cite{GBCKL12}. Further developments may include the implementation of higher-dimensionality KS sets \cite{KP95} and portable KS sets in integrated photonic circuits \cite{PCRYO08,PLMMPPZLIWBSTO10,SSVMCRO10,SSVMCRO12}. Other developments that could be pursued in future work are device-independent security that is based on contextuality \cite{Cabello10,HHHHPB10} and state-independent quantum correlations with computational power \cite{AB09}.


\begin{acknowledgments}
We acknowledge fruitful discussions with A.~J.~L\'opez-Tarrida and thank M.~R{\aa }dmark for technical support with the experiment. This work was supported by the Spanish Ministry of Economy and Competitiveness Project No.~FIS2011-29400, the Wenner-Gren Foundation, FIRB Futuro in Ricerca-HYTEQ, the Swedish Research Council (VR), the Linnaeus Center of Excellence ADOPT, ERC Advanced Grant QOLAPS, and Project PHORBITECH of the Future and Emerging Technologies (FET) Program within the Seventh Framework Programme for Research of the European Commission, under FET-Open Grant No.~255914.
\end{acknowledgments}


\appendix


\section{Proof that the set of classical tests in Fig.~1(b) provides the maximum probability of obtaining yes using classical resources}
\label{AppA}


The maximum probability of obtaining a result~1 (yes) when a yes-no test is chosen at random and using classical resources is given by $\alpha(G)/n(G)$, where $G$ is the graph defined in the main text, $\alpha(G)$ is its independence number (defined as the maximum number of pairwise nonadjacent vertices in $G$), and $n(G)$ is the number of vertices of $G$ \cite{CSW10}. For the graph in Fig.~1(b), $\alpha(G)=4$ and $n(G)=18$. For the set of classical tests in Fig.~1(b), the probability is $4/18$, no matter in which box the ball was initially placed. On the other hand, the minimum number of classical states needed to accomplish the task is given by the intersection number of the complement of $G$, $\theta'(\bar{G})$ (defined as the smallest number of subsets of pairwise adjacent vertices needed to cover all of the edges of the complement of $G$). For the graph in Fig.~1(b), $\theta'(\bar{G})=18$, which shows that the strategy in Fig.~1(b) is also optimal in the sense that it uses the smallest possible classical system.\hfill \endproof


\section{Proof that the highest probability for the task with quantum resources cannot be outperformed using post-quantum resources}
\label{AppB}


The maximum probability of obtaining a result~1 (yes) when a yes-no test is chosen at random and using post-quantum resources is given by $\alpha(G)^*/n(G)$, where $\alpha^*(G)$ is the fractional packing number of $G$, which is defined as
\begin{equation}
 \alpha^*(G) = \max \sum_{i\in V(G)} w_i,
\end{equation}
where $V(G)$ is the set of vertices of $G$, and the maximum is taken for all $0 \leq w_i\leq 1$ and for all subsets of pairwise adjacent vertices $c_j$ of $G$, under the restriction $\sum_{i \in c_j} w_i \leq 1$ \cite{CSW10}. For the graph in Fig.~1(b), $\alpha^*(G)=4.5$.\hfill \endproof


\section{Proof that the highest quantum violation of the noncontextuality inequality (5) cannot be outperformed using post-quantum resources}
\label{AppC}


The highest quantum violation of a noncontextuality inequality which can be expressed as a sum of joint probabilities $P(abc|xyz)$ is given by the Lov\'asz number $\vartheta^(G)$ of the graph in which each proposition $abc|xyz$ is represented by a vertex and exclusive propositions correspond to adjacent vertices \cite{CSW10}. The Lov\'asz number of a graph $G$ is
\begin{equation}\label{theta}
\vartheta(G)=\max \sum_{i \in V(G)} |\langle\psi|v_{i}\rangle|^{2},
\end{equation}
where the maximum is taken over all unit vectors $|\psi\rangle$ and $|v_{i}\rangle$ and all dimensions, where each $|v_{i}\rangle$ corresponds to a vertex of $G$, and two vertices are adjacent if and only if the corresponding vectors are orthogonal.

On the other hand, the highest violation of a noncontextuality inequality satisfying that the sum of the probabilities of pairwise exclusive events cannot be higher than~1 is given by $\alpha^*(G)$ \cite{CSW10}.

The graph $G$ for the noncontextuality inequality (5) is the one in Fig.~1(c), which has $\vartheta(G)=\alpha^*(G)=4.5$.\hfill \endproof




\begin{table}[htb]
{\small
\begin{tabular}{|c|c|c|c|}
\hline
\hline
\textbf{Code} & \textbf{State} & \textbf{Implementation 1} & \textbf{Implementation 2} \\
\hline\hline
$v_{1}$ & $(1,0,0,0)$ & $\ket{H, +2}$ & $\ket{t, H}$ \\ \hline
$v_{2}$ & $(0,1,0,0)$ & $\ket{H, -2}$ & $\ket{t, V}$ \\ \hline
$v_{3}$ & $(0,0,1,1)$ & $\ket{V, h}$ & $\ket{r, A}$ \\ \hline
$v_{4}$ & $(0,0,1,-1)$ & $\ket{V, v}$ & $\ket{r, D}$ \\ \hline
$v_{5}$ & $(1,-1,0,0)$ & $\ket{H, v}$ & $\ket{t, D}$ \\ \hline
$v_{6}$ & $(1,1,-1,-1)$ & $\ket{D, h}$ & $\ket{s, A}$ \\ \hline
$v_{7}$ & $(1,1,1,1)$ & $\ket{A, h}$ & $\ket{p, A}$ \\ \hline
$v_{8}$ & $(1,-1,1,-1)$ & $\ket{A, v}$ & $\ket{p, D}$ \\ \hline
$v_{9}$ & $(1,0,-1,0)$ & $\ket{D, +2}$ & $\ket{s, H}$ \\ \hline
$v_{10}$ & $(0,1,0,-1)$ & $\ket{D, -2}$ & $\ket{s, V}$ \\ \hline
$v_{11}$ & $(1,0,1,0)$ & $\ket{A, +2}$ & $\ket{p, H}$ \\ \hline
$v_{12}$ & $(1,1,-1,1)$ & $\frac{1}{\sqrt{2}}(\ket{D, +2}+\ket{A, -2})$ & $\frac{1}{\sqrt{2}}(\ket{t, A}-\ket{r, D})$ \\ \hline
$v_{13}$ & $(-1,1,1,1)$ & $\frac{1}{\sqrt{2}}(\ket{A,-2}-\ket{D,+2})$ & $\frac{1}{\sqrt{2}}(\ket{t, D}-\ket{r, A})$ \\ \hline
$v_{14}$ & $(1,1,1,-1)$ & $\frac{1}{\sqrt{2}}(\ket{A,+2}+\ket{D,-2})$ & $\frac{1}{\sqrt{2}}(\ket{t, A}+\ket{r, D})$ \\ \hline
$v_{15}$ & $(1,0,0,1)$ & $\frac{1}{\sqrt{2}}(\ket{H,+2}+\ket{V,-2})$ & $\frac{1}{\sqrt{2}}(\ket{t, H}+\ket{r, V})$ \\ \hline
$v_{16}$ & $(0,1,-1,0)$ & $\frac{1}{\sqrt{2}}(\ket{H,-2}-\ket{V,+2})$ & $\frac{1}{\sqrt{2}}(\ket{r, H}-\ket{t, V})$ \\ \hline
$v_{17}$ & $(0,1,1,0)$ & $\frac{1}{\sqrt{2}}(\ket{H,-2}+\ket{V,+2})$ & $\frac{1}{\sqrt{2}}(\ket{r, H}+\ket{t, V})$ \\ \hline
$v_{18}$ & $(0,0,0,1)$ & $\ket{V,-2}$ & $\ket{r, V}$ \\ \hline
 \hline
\end{tabular}
}
\caption{Implementations of 18 eigenstates of the KS set used in the experiments 1 and 2. Notation for this Table includes
$\ket{A} = \frac{\ket{H}+\ket{V}}{\sqrt{2}}$,
$\ket{D} = \frac{\ket{H}-\ket{V}}{\sqrt{2}}$,
$\ket{h} = \frac{\ket{+2}+\ket{-2}}{\sqrt{2}}$,
$\ket{v} = \frac{\ket{+2}-\ket{-2}}{\sqrt{2}}$,
$\ket{p} = \frac{\ket{t}+\ket{r}}{\sqrt{2}}$, and
$\ket{s} = \frac{\ket{t}-\ket{r}}{\sqrt{2}}$.
Each state $v_i$ corresponds to a vertex in the graph in Fig.~1(a).}
\label{prob}
\end{table}


\begin{table}[htb]
{\small
\begin{tabular}{|c|c|c|}
\hline
\hline
\textbf{State} & \textbf{Implementation 1} & $\Sigma$ \\
\hline\hline
$v_{1}$ & $\ket{H,+2}$ & 4.60 \\ \hline
$v_{2}$ & $\ket{H,-2}$ & 4.45 \\ \hline
$v_{3}$ & $\ket{V,h}$ & 4.51 \\ \hline
$v_{4}$ & $\ket{V,v}$ & 4.63 \\ \hline
$v_{5}$ & $\ket{H,v}$ & 4.54\\ \hline
$v_{6}$ & $\ket{D,h}$ & 4.55\\ \hline
$v_{7}$ & $\ket{A,h}$ & 4.50\\ \hline
$v_{8}$ & $\ket{A,v}$ & 4.45\\ \hline
$v_{9}$ & $\ket{D,+2}$ & 4.48\\ \hline
$v_{10}$ & $\ket{D,-2}$ & 4.65\\ \hline
$v_{11}$ & $\ket{A,+2}$ & 4.51\\ \hline
$v_{12}$ & $\frac{1}{\sqrt{2}}(\ket{D,+2}+\ket{A,-2})$ & 4.42\\ \hline
$v_{13}$ & $\frac{1}{\sqrt{2}}(\ket{A,-2}-\ket{D,+2})$ & 4.51\\ \hline
$v_{14}$ & $\frac{1}{\sqrt{2}}(\ket{A,+2}+\ket{D,-2})$ & 4.52\\ \hline
$v_{15}$ & $\ket{\psi_1}=\frac{1}{\sqrt{2}}(\ket{H,+2}+\ket{V,-2})$ & 4.59\\ \hline
$v_{16}$ & $\ket{\psi_3}=\frac{1}{\sqrt{2}}(\ket{H,-2}-\ket{V,+2})$ & 4.47\\ \hline
$v_{17}$ & $\ket{\psi_4}=\frac{1}{\sqrt{2}}(\ket{H,-2}+\ket{V,+2})$ & 4.41\\ \hline
$v_{18}$ & $\ket{V,-2}$ & 4.50\\ \hline
$v_{19}$ & $\ket{V,+2}$ & 4.45\\ \hline
$v_{20}$ & $\ket{H,h}$ & 4.57\\ \hline
$v_{21}$ & $\ket{A,-2}$ & 4.37\\ \hline
$v_{22}$ & $\ket{D,v}$ & 4.45\\ \hline
$v_{23}$ & $\frac{1}{\sqrt{2}}(\ket{A,+2}-\ket{D,-2})$ & 4.42\\ \hline
$v_{24}$ & $\ket{\psi_2}=\frac{1}{\sqrt{2}}(\ket{H,+2}-\ket{V,-2})$ & 4.58\\ \hline
$\rho_{25}$ & $\frac{13}{16}\ket{\psi_1}\bra{\psi_1}+\frac{1}{16}\sum_{j=2}^4\ket{\psi_j}\bra{\psi_j}$ & 4.57\\ \hline
$\rho_{26}$ & $\frac{5}{8}\ket{\psi_1}\bra{\psi_1}+\frac{1}{8}\sum_{j=2}^4\ket{\psi_j}\bra{\psi_j}$ & 4.55\\ \hline
$\rho_{27}$ & $\frac{7}{16}\ket{\psi_1}\bra{\psi_1}+\frac{3}{16}\sum_{j=2}^4\ket{\psi_j}\bra{\psi_j}$ & 4.53\\ \hline
$\rho_{28}$ & $\frac{1}{4}\sum_{j=1}^4\ket{\psi_j}\bra{\psi_j}$ & 4.50\\ \hline
 \hline
 \multicolumn{2}{|c|}{\textbf{Average value}} & \textbf{$ 4.51 $}\\
\hline
\end{tabular}
}
\caption{Experimental results for $\Sigma$, defined in (1), for 28 quantum states. Notation for this Table includes
$\ket{A} = \frac{\ket{H}+\ket{V}}{\sqrt{2}}$,
$\ket{D} = \frac{\ket{H}-\ket{V}}{\sqrt{2}}$,
$\ket{h} = \frac{\ket{+2}+\ket{-2}}{\sqrt{2}}$, and
$\ket{v} = \frac{\ket{+2}-\ket{-2}}{\sqrt{2}}$.}
\label{prob}
\end{table}


\begin{table}[htb]
{\small
\begin{tabular}{|c|c|c|c|c|c|}
\hline\hline
\multicolumn{2}{|c|}{\textbf{Probabilities}} & \multicolumn{2}{|c|}{\textbf{Probabilities}}& \multicolumn{2}{|c|}{\textbf{Probabilities}}\\
\hline
$p_{i,j}$ & Value & $p_{j,i}$ & Value &$p_{i,j}$ & Value \\
\hline\hline
$p_{1,2}$ & 0.02 &$p_{7,4}$ & 0.02 &$p_{13,4}$ & 0.01 \\
$p_{1,3}$ & 0 &$p_{7,5}$ & 0.02 &$p_{13,10}$ & 0.002 \\
$p_{1,4}$ & 0 &$p_{7,6}$ & 0 &$p_{13,11}$ & 0.006 \\
$p_{1,10}$ & 0.008 &$p_{7,8}$ & 0.04 &$p_{13,12}$ & 0.006 \\
$p_{1,16}$ & 0.02 &$p_{7,9}$ & 0 &$p_{13,14}$ & 0.064 \\
$p_{1,17}$ & 0.006 &$p_{7,10}$ & 0 &$p_{13,15}$ & 0.032 \\
$p_{1,18}$ & 0 &$p_{7,16}$ & 0.003 &$p_{13,16}$ & 0.08 \\

$p_{2,1}$ & 0.003 & $p_{8,3}$ & 0.017 &$p_{14,3}$ & 0.035 \\
$p_{2,3}$ & 0 & $p_{8,6}$ & 0 &$p_{14,5}$ & 0.018 \\
$p_{2,4}$ & 0 &$p_{8,7}$ & 0.035 &$p_{14,9}$ & 0.009 \\
$p_{2,9}$ & 0 &$p_{8,7}$ & 0 &$p_{14,12}$ & 0.005 \\
$p_{2,11}$ & 0.01 &$p_{8,10}$ & 0&$p_{14,13}$ & 0.004 \\
$p_{2,15}$ & 0.007 &$p_{8,15}$ & 0.03 &$p_{14,15}$ & 0.02 \\
$p_{2,18}$ & 0 &$p_{8,17}$ & 0 &$p_{14,16}$ & 0.012 \\

$p_{3,1}$ & 0 & $p_{9,2}$ & 0.014 &$p_{15,2}$ & 0.029 \\
$p_{3,2}$ & 0 &$p_{9,7}$ & 0 &$p_{15,6}$ & 0.01 \\
$p_{3,4}$ & 0.04 &$p_{9,8}$ & 0 &$p_{15,8}$ & 0.03 \\
$p_{3,5}$ & 0 &$p_{9,10}$ & 0.034 &$p_{15,13}$ & 0.02 \\
$p_{3,8}$ & 0.015 &$p_{9,11}$ & 0 &$p_{15,14}$ & 0.01 \\
$p_{3,12}$ & 0.003 &$p_{9,14}$ & 0.013 &$p_{15,16}$ & 0.01 \\
$p_{3,14}$ & 0.02 &$p_{9,18}$ & 0.01 &$p_{15,17}$ & 0.01 \\

$p_{4,1}$ & 0 &$p_{10,1}$ & 0 &$p_{16,1}$ & 0.01 \\
$p_{4,2}$ & 0 &$p_{10,7}$ & 0 &$p_{16,7}$ & 0.017\\
$p_{4,3}$ & 0.042 &$p_{10,8}$ & 0 &$p_{16,13}$ & 0.01 \\
$p_{4,5}$ & 0 &$p_{10,9}$ & 0.009 &$p_{16,14}$ & 0.06 \\
$p_{4,6}$ & 0.02 & $p_{10,11}$ & 0 &$p_{16,15}$ & 0.005 \\
$p_{4,7}$ & 0.03 &$p_{10,12}$ & 0.032 &$p_{16,17}$ & 0.03 \\
$p_{4,13}$ & 0 & $p_{10,13}$ & 0.053&$p_{16,18}$ & 0.02 \\

$p_{5,3}$ & 0 &$p_{11,2}$ & 0.01 &$p_{17,1}$ & 0.007 \\
$p_{5,4}$ & 0 &$p_{11,6}$ & 0 &$p_{17,6}$ & 0.005 \\
$p_{5,6}$ & 0.02 &$p_{11,9}$ & 0 &$p_{17,8}$ & 0.02\\
$p_{5,7}$ & 0.02 &$p_{11,10}$ & 0 &$p_{17,12}$ & 0.022 \\
$p_{5,12}$ & 0.02 & $p_{11,12}$ & 0.006 &$p_{17,15}$ & 0.04 \\
$p_{5,14}$ & 0.006 &$p_{11,13}$ & 0.003 &$p_{17,16}$ & 0.03 \\
$p_{5,18}$ & 0 & $p_{11,18}$ & 0.01 &$p_{17,18}$ & 0.03 \\

$p_{6,4}$ & 0.01 &$p_{12,3}$ & 0.009 &$p_{18,1}$ & 0 \\
$p_{6,5}$ & 0.02 &$p_{12,5}$ & 0.028 &$p_{18,2}$ & 0 \\
$p_{6,7}$ & 0 &$p_{12,10}$ & 0.033 &$p_{18,5}$ & 0 \\
$p_{6,8}$ & 0 &$p_{12,11}$ & 0.017 &$p_{18,9}$ & 0 \\
$p_{6,11}$ & 0 &$p_{12,13}$ & 0.031 & $p_{18,11}$ & 0.001 \\
$p_{6,15}$ & 0.04 &$p_{12,14}$ & 0 &$p_{18,16}$ & 0.016 \\
$p_{6,17}$ & 0.09 &$p_{12,17}$ & 0.069 &$p_{18,17}$ & 0.038 \\
\hline
\end{tabular}
}
\caption{Test of the exclusivity relations. Experimental values for the probabilities $P_{|v_j\rangle}(\Pi_i=1)=p_{i,j}$ used for the demonstration of the exclusivity relations between the KS tests. Each pair $\{i,j\}$ corresponds to an edge in the graph in Fig.~1(a).}
\label{prob}
\end{table}




\begin{table*}[h]
{\small
\begin{tabular}{|c|c|c|c|c|c|c|}
\hline
\hline
States & $P(001|012)$ & $P(111|012)$ & $P(100|012)$ & $P(010|036)$ & $P(001|036)$ & $P(100|036)$ \\
\hline\hline
 $v_{1}$ & 0.00469 $\pm$ 0.00003 & 0.96943 $\pm$ 0.00008 & 0.00235 $\pm$ 0.00002 & 0.00688 $\pm$ 0.00003 & 0.00246 $\pm$ 0.00002 & 0.46080 $\pm$ 0.00039 \\
 $v_{2}$ & 0.00816 $\pm$ 0.00006 & 0.00616 $\pm$ 0.00005 & 0.95422 $\pm$ 0.00015 & 0.00831 $\pm$ 0.00005 & 0.00331 $\pm$ 0.00003 & 0.46099 $\pm$ 0.00190 \\
 $v_{7}$ & 0.25364 $\pm$ 0.00048 & 0.29421 $\pm$ 0.00057 & 0.24430 $\pm$ 0.00047 & 0.40141 $\pm$ 0.00013 & 0.06211 $\pm$ 0.00007 & 0.09906 $\pm$ 0.00007 \\
$v_{11}$ & 0.01102 $\pm$ 0.00007 & 0.59332 $\pm$ 0.00195 & 0.00507 $\pm$ 0.00004 & 0.23417 $\pm$ 0.00032 & 0.23857 $\pm$ 0.00036 & 0.23913 $\pm$ 0.00029 \\
$v_{15}$ & 0.47407 $\pm$ 0.00034 & 0.51573 $\pm$ 0.00034 & 0.00079 $\pm$ 0.00001 & 0.17226 $\pm$ 0.00014 & 0.28003 $\pm$ 0.00020 & 0.24091 $\pm$ 0.00021 \\
$v_{16}$ & 0.00864 $\pm$ 0.00005 & 0.00579 $\pm$ 0.00005 & 0.50176 $\pm$ 0.00131 & 0.23686 $\pm$ 0.00071 & 0.23889 $\pm$ 0.00077 & 0.24422 $\pm$ 0.00075 \\
$v_{17}$ & 0.01085 $\pm$ 0.00007 & 0.01010 $\pm$ 0.00007 & 0.49884 $\pm$ 0.00173 & 0.23263 $\pm$ 0.00063 & 0.23713 $\pm$ 0.00066 & 0.24524 $\pm$ 0.00063 \\
$v_{18}$ & 0.95668 $\pm$ 0.00062 & 0.00303 $\pm$ 0.00006 & 0.00477 $\pm$ 0.00008 & 0.41078 $\pm$ 0.00086 & 0.57672 $\pm$ 0.00087 & 0.00077 $\pm$ 0.00001 \\
$v_{19}$ & 0.02200 $\pm$ 0.00017 & 0.01389 $\pm$ 0.00013 & 0.00482 $\pm$ 0.00006 & 0.48524 $\pm$ 0.00194 & 0.49788 $\pm$ 0.00193 & 0.00157 $\pm$ 0.00002 \\
$v_{20}$ & 0.01466 $\pm$ 0.00008 & 0.47650 $\pm$ 0.00068 & 0.46002 $\pm$ 0.00068 & 0.00974 $\pm$ 0.00004 & 0.00269 $\pm$ 0.00002 & 0.18468 $\pm$ 0.00044 \\
$v_{24}$ & 0.47056 $\pm$ 0.00100 & 0.50962 $\pm$ 0.00100 & 0.00139 $\pm$ 0.00001 & 0.23378 $\pm$ 0.00074 & 0.28746 $\pm$ 0.00086 & 0.23480 $\pm$ 0.00077 \\
$\rho_{25}$ & 0.45592 $\pm$ 0.01628 & 0.49575 $\pm$ 0.01645 & 0.02004 $\pm$ 0.00694 & 0.18569 $\pm$ 0.00228 & 0.27469 $\pm$ 0.00246 & 0.24079 $\pm$ 0.00142 \\
$\rho_{26}$ & 0.42127 $\pm$ 0.02612 & 0.45821 $\pm$ 0.02656 & 0.05688 $\pm$ 0.01134 & 0.19367 $\pm$ 0.00264 & 0.27209 $\pm$ 0.00290 & 0.24065 $\pm$ 0.00164 \\
$\rho_{27}$ & 0.38695 $\pm$ 0.03270 & 0.42054 $\pm$ 0.03347 & 0.09327 $\pm$ 0.01447 & 0.20662 $\pm$ 0.00274 & 0.26551 $\pm$ 0.00335 & 0.24091 $\pm$ 0.00178 \\
$\rho_{28}$ & 0.37036 $\pm$ 0.03452 & 0.40131 $\pm$ 0.03542 & 0.11119 $\pm$ 0.01547 & 0.21979 $\pm$ 0.00252 & 0.26237 $\pm$ 0.00339 & 0.24033 $\pm$ 0.00166 \\
\hline
\end{tabular}
}
\caption{Experimental values of 6 out of the 18 probabilities needed to test the violation of the noncontextuality inequality (5), for 15 states. For the other 12 probabilities, see Tables \ref{table_2} and \ref{table_3}.
These experimental values lead to the values of $\xi$ reported in Table~II.}
\label{table_1}
\end{table*}


\begin{table*}[h]
{\small
\begin{tabular}{|c|c|c|c|c|c|c|}
\hline
\hline
States & $P(100|345)$ & $P(111|345)$ & $P(010|345)$ & $P(100|147)$ & $P(001|147)$ & $P(111|147)$ \\
\hline\hline
 $v_{1}$& 0.26452 $\pm$ 0.00024 & 0.20060 $\pm$ 0.00020 & 0.27066 $\pm$ 0.00026 & 0.53399 $\pm$ 0.00027 & 0.01821 $\pm$ 0.00007 & 0.38350 $\pm$ 0.00025 \\
 $v_{2}$& 0.22402 $\pm$ 0.00017 & 0.21884 $\pm$ 0.00018 & 0.29808 $\pm$ 0.00022 & 0.02381 $\pm$ 0.00004 & 0.39155 $\pm$ 0.00020 & 0.02196 $\pm$ 0.00004 \\
 $v_{7}$& 0.00447 $\pm$ 0.00002 & 0.91502 $\pm$ 0.00028 & 0.01106 $\pm$ 0.00003 & 0.00967 $\pm$ 0.00010 & 0.02254 $\pm$ 0.00013 & 0.47878 $\pm$ 0.00021 \\
$v_{11}$& 0.02834 $\pm$ 0.00012 & 0.47805 $\pm$ 0.00028 & 0.44096 $\pm$ 0.00028 & 0.02304 $\pm$ 0.00023 & 0.08027 $\pm$ 0.00034 & 0.84749 $\pm$ 0.00041 \\
$v_{15}$& 0.00139 $\pm$ 0.00002 & 0.47787 $\pm$ 0.00014 & 0.00158 $\pm$ 0.00002 & 0.24296 $\pm$ 0.00051 & 0.23698 $\pm$ 0.00040 & 0.25704 $\pm$ 0.00052 \\
$v_{16}$& 0.39052 $\pm$ 0.00020 & 0.01284 $\pm$ 0.00007 & 0.53917 $\pm$ 0.00020 & 0.21978 $\pm$ 0.00023 & 0.27539 $\pm$ 0.00036 & 0.19372 $\pm$ 0.00024 \\
$v_{17}$& 0.01329 $\pm$ 0.00007 & 0.45783 $\pm$ 0.00021 & 0.01498 $\pm$ 0.00008 & 0.22005 $\pm$ 0.00015 & 0.25323 $\pm$ 0.00024 & 0.22106 $\pm$ 0.00016 \\
$v_{18}$& 0.28602 $\pm$ 0.00016 & 0.25611 $\pm$ 0.00016 & 0.21268 $\pm$ 0.00020 & 0.03352 $\pm$ 0.00014 & 0.40537 $\pm$ 0.00041 & 0.03545 $\pm$ 0.00012 \\
$v_{19}$& 0.22260 $\pm$ 0.00018 & 0.27739 $\pm$ 0.00020 & 0.24072 $\pm$ 0.00018 & 0.50666 $\pm$ 0.00029 & 0.01810 $\pm$ 0.00005 & 0.41785 $\pm$ 0.00028 \\
$v_{20}$& 0.49902 $\pm$ 0.00031 & 0.43034 $\pm$ 0.00028 & 0.02680 $\pm$ 0.00027 & 0.33736 $\pm$ 0.00019 & 0.14444 $\pm$ 0.00014 & 0.25529 $\pm$ 0.00015 \\
$v_{24}$& 0.51553 $\pm$ 0.00026 & 0.00638 $\pm$ 0.00008 & 0.44622 $\pm$ 0.00026 & 0.25113 $\pm$ 0.00038 & 0.16467 $\pm$ 0.00040 & 0.28708 $\pm$ 0.00043 \\
$\rho_{25}$& 0.07222 $\pm$ 0.02157 & 0.40495 $\pm$ 0.02121 & 0.07689 $\pm$ 0.02201 & 0.23925 $\pm$ 0.00136 & 0.23636 $\pm$ 0.00290 & 0.25100 $\pm$ 0.00204 \\
$\rho_{26}$& 0.12609 $\pm$ 0.02610 & 0.34961 $\pm$ 0.02528 & 0.13397 $\pm$ 0.02638 & 0.23742 $\pm$ 0.00156 & 0.23441 $\pm$ 0.00357 & 0.24837 $\pm$ 0.00241 \\
$\rho_{27}$& 0.18242 $\pm$ 0.02788 & 0.29159 $\pm$ 0.02634 & 0.19362 $\pm$ 0.02793 & 0.23597 $\pm$ 0.00181 & 0.22921 $\pm$ 0.00445 & 0.24599 $\pm$ 0.00286 \\
$\rho_{28}$& 0.24562 $\pm$ 0.02897 & 0.22560 $\pm$ 0.02666 & 0.26257 $\pm$ 0.02892 & 0.23366 $\pm$ 0.00183 & 0.23354 $\pm$ 0.00458 & 0.23879 $\pm$ 0.00297 \\
\hline
\end{tabular}
}
\caption{Experimental values of 6 out of the 18 probabilities needed to test the violation of the noncontextuality inequality (5), text for 15 states. For the other 12 probabilities, see Tables \ref{table_1} and \ref{table_3}.
These experimental values lead to the values of $\xi$ reported in Table~II.}
\label{table_2}
\end{table*}


\begin{table*}[h]
{\small
\begin{tabular}{|c|c|c|c|c|c|c|}
\hline
\hline
States & $P(100|678)$ & $P(001|678)$ & $P(111|678)$ & $P(110|258)$ & $P(000|258)$ & $P(011|258)$ \\
\hline\hline
 $v_{1}$& 0.20429 $\pm$ 0.00026 & 0.20198 $\pm$ 0.00036 & 0.21255 $\pm$ 0.00022 & 0.45072 $\pm$ 0.00124 & 0.00162 $\pm$ 0.00001 & 0.00606 $\pm$ 0.00003 \\
 $v_{2}$& 0.21116 $\pm$ 0.00019 & 0.27606 $\pm$ 0.00028 & 0.20309 $\pm$ 0.00030 & 0.00370 $\pm$ 0.00002 & 0.46596 $\pm$ 0.00107 & 0.48961 $\pm$ 0.00108 \\
 $v_{7}$& 0.24464 $\pm$ 0.00027 & 0.17138 $\pm$ 0.00013 & 0.24261 $\pm$ 0.00026 & 0.34784 $\pm$ 0.00028 & 0.01168 $\pm$ 0.00005 & 0.56462 $\pm$ 0.00036 \\
$v_{11}$& 0.00696 $\pm$ 0.00005 & 0.00827 $\pm$ 0.00006 & 0.45328 $\pm$ 0.00044 & 0.22640 $\pm$ 0.00053 & 0.21397 $\pm$ 0.00053 & 0.31228 $\pm$ 0.00073 \\
$v_{15}$& 0.40428 $\pm$ 0.00013 & 0.01273 $\pm$ 0.00002 & 0.00945 $\pm$ 0.00003 & 0.92373 $\pm$ 0.00033 & 0.00439 $\pm$ 0.00002 & 0.00448 $\pm$ 0.00002 \\
$v_{16}$& 0.43543 $\pm$ 0.00016 & 0.01655 $\pm$ 0.00009 & 0.00642 $\pm$ 0.00002 & 0.01271 $\pm$ 0.00002 & 0.88871 $\pm$ 0.00037 & 0.02764 $\pm$ 0.00009 \\
$v_{17}$& 0.00580 $\pm$ 0.00002 & 0.47776 $\pm$ 0.00042 & 0.39824 $\pm$ 0.00050 & 0.00173 $\pm$ 0.00001 & 0.02191 $\pm$ 0.00003 & 0.87828 $\pm$ 0.00031 \\
$v_{18}$& 0.16293 $\pm$ 0.00012 & 0.28624 $\pm$ 0.00020 & 0.17858 $\pm$ 0.00032 & 0.48059 $\pm$ 0.00030 & 0.00743 $\pm$ 0.00002 & 0.00240 $\pm$ 0.00001 \\
$v_{19}$& 0.23814 $\pm$ 0.00020 & 0.22213 $\pm$ 0.00015 & 0.21773 $\pm$ 0.00018 & 0.00918 $\pm$ 0.00003 & 0.40457 $\pm$ 0.00077 & 0.53408 $\pm$ 0.00078 \\
$v_{20}$& 0.38909 $\pm$ 0.00030 & 0.04651 $\pm$ 0.00006 & 0.40079 $\pm$ 0.00032 & 0.18890 $\pm$ 0.00016 & 0.26179 $\pm$ 0.00019 & 0.28265 $\pm$ 0.00021 \\
$v_{24}$& 0.01863 $\pm$ 0.00005 & 0.46692 $\pm$ 0.00023 & 0.32780 $\pm$ 0.00026 & 0.01507 $\pm$ 0.00008 & 0.00529 $\pm$ 0.00001 & 0.00447 $\pm$ 0.00001 \\
$\rho_{25}$& 0.34260 $\pm$ 0.01553 & 0.08828 $\pm$ 0.01954 & 0.06934 $\pm$ 0.01618 & 0.70423 $\pm$ 0.04253 & 0.06367 $\pm$ 0.02803 & 0.09191 $\pm$ 0.03489 \\
$\rho_{26}$& 0.29852 $\pm$ 0.01854 & 0.14204 $\pm$ 0.02265 & 0.10917 $\pm$ 0.01866 & 0.57473 $\pm$ 0.04545 & 0.13755 $\pm$ 0.03791 & 0.11320 $\pm$ 0.03737 \\
$\rho_{27}$& 0.26977 $\pm$ 0.01972 & 0.17678 $\pm$ 0.02360 & 0.13295 $\pm$ 0.01927 & 0.32880 $\pm$ 0.04335 & 0.18683 $\pm$ 0.04094 & 0.22940 $\pm$ 0.04564 \\
$\rho_{28}$& 0.22907 $\pm$ 0.02072 & 0.22915 $\pm$ 0.02352 & 0.17698 $\pm$ 0.02009 & 0.21618 $\pm$ 0.03903 & 0.25357 $\pm$ 0.04370 & 0.24670 $\pm$ 0.04584 \\
\hline
\end{tabular}
}
\caption{Experimental values of 6 out of the 18 probabilities needed to test the violation of the noncontextuality inequality (5), for 15 states. For the other 12 probabilities, see Tables \ref{table_1} and \ref{table_2}.
These experimental values lead to the values of $\xi$ reported in Table~II.}
\label{table_3}
\end{table*}



\end{document}